**Attosecond state-resolved carrier motion in quantum materials probed by soft X-ray XANES**

Bárbara Buades[1], Antonio Picón[1,2], Emma Berger[3,4], Iker León[1], Nicola Di Palo[1], Seth L. Cousin[1], Caterina Cocchi[5], Eric Pellegrin[6], Javier Herrero Martin[6], Samuel Mañas-Valero[7], Eugenio Coronado[7], Thomas Danz[8], Claudia Draxl[5], Mitsuharu Uemoto[9], Kazuhiro Yabana[9], Martin Schultze[10], Simon Wall[1], Michael Zürch[3,4,11], Jens Biegert[1,11,12,*]

*[1]ICFO-Institut de Ciencies Fotoniques, The Barcelona Institute of Science and Technology, Castelldefels (Barcelona), Spain*
*[2]Departamento de Química, Universidad Autónoma de Madrid, Madrid, Spain*
*[3]University of California at Berkeley, Department of Chemistry, Berkeley, CA, USA*
*[4]Lawrence Berkeley National Laboratory, Materials Sciences Division, Berkeley, CA, USA*
*[5]Institut für Physik and IRIS Adlershof, Humboldt-Universität zu Berlin, Berlin, Germany*
*[6]ALBA Synchrotron Light Source, Cerdanyola del Vallès, Barcelona, Spain*
*[7]Instituto de Ciencia Molecular (ICMol), Universitat de València, Paterna, Spain*
*[8]4th Physical Institute - Solids and Nanostructures, University of Göttingen, Göttingen, Germany.*
*[9]Center for Computational Sciences, University of Tsukuba, Tsukuba, Japan*
*[10]Institute of Experimental Physics, Graz University of Technology, Graz, Austria*
*[11]Fritz Haber Institute of the Max Planck Society, Berlin, Germany*
*[12]ICREA, Pg. Lluís Companys 23, Barcelona, Spain*
**Correspondence to: jens.biegert@icfo.eu*

## Abstract

Recent developments in attosecond technology led to tabletop X-ray spectroscopy in the soft X-ray range, thus uniting the element- and state-specificity of core-level x-ray absorption spectroscopy with the time resolution to follow electronic dynamics in real time. We describe recent work in attosecond technology and investigations into materials such as Si, $SiO_2$, GaN, $Al_2O_3$, Ti, $TiO_2$, enabled by the convergence of these two capabilities. We showcase the state-of-the-art on isolated attosecond soft x-ray pulses for x-ray absorption near edge spectroscopy (XANES) to observe the 3d-state dynamics of the semi-metal $TiS_2$ with attosecond resolution at the Ti L-edge (460 eV). We describe how the element- and state-specificity at the transition metal L-edge of the quantum material allows to unambiguously identify how and where the optical field influences charge carriers. This precision elucidates that the Ti:3d conduction band states are efficiently photo-doped to a density of $1.9 \times 10^{21}$ $cm^{-3}$ and that the light-field induces coherent motion of intra-band carriers across 38% of the first Brillouin zone. Lastly, we describe the prospects with such unambiguous real-time observation of carrier dynamics in specific bonding or anti-bonding states and speculate that such capability will bring unprecedented opportunities towards an engineered approach for designer materials with pre-defined properties and efficiency. Examples are composites of semiconductors and insulators like Si, Ge, $SiO_2$, GaN, BN, quantum materials like graphene, TMDCs, or high-Tc superconductors like NbN or LaBaCuO. Exiting are prospects to scrutinize canonical questions in multi-body physics such as whether the electrons or lattice trigger phase transitions.

## I. Introduction

Until the last decade, ultrafast soft x-ray spectroscopies were nearly exclusively available at synchrotrons, where tens- to hundreds-of-picoseconds temporal resolution was “state-of-the-art”. To improve time resolution for time-resolved experiments, femtosecond slicing was developed to improve temporal resolution to the ~150 femtoseconds range[1,2], but on the expense of photon flux. Alternatively, table-top laser-driven plasma sources could provide incoherent emission of soft and hard X-ray bursts of radiation with temporal duration in the sub-100 femtosecond range[3,4], but with the difficulty of needing shielding from debris and the requirement for high numerical aperture x-ray optics to capture the incoherent x-ray emission. With the discovery of high-harmonic generation (HHG)[5,6], the first table-top technology became available, which yields fully coherent radiation and achieves attosecond temporal resolution. However, HHG suffered from low flux and low photon energies in the XUV range, which precluded its utility for core-level X-ray spectroscopy since the conditions to achieve element- and state-specificity could not be reached. However, during the last years, several breakthroughs were achieved to leverage the prospects of ponderomotive scaling of HHG[7,8]: Isolated attosecond soft X-ray pulses[9–12] with water-window coverage[9,13–15] are nowadays available, and the utility of attosecond technology for XANES[9,16,17] and EXAFS[18,19] has been demonstrated. In parallel with these developments, the first self-amplified spontaneous emission (SASE) operation of an accelerator has been shown to yield x-ray radiation[20]. This has led to the development of so-called "4$^{th}$ generation light sources" in the accelerator community, namely large-scale x-ray free electron laser facilities (XFELs), which provide unsurpassed photon flux (mostly) at hard x-ray photon energies. Due to the inherent randomness of the SASE process, intensive developments are under way to achieve reproducible pulse duration and temporal synchronization. Presently, there exist only a handful of XFELs worldwide, each having distinct parameters, with only very recently the demonstrated capability of delivering sub-femtosecond pulses with good coherence and sufficiently high repetition rates[21]. Further, performing experiments at such billion-euro-scale facilities poses its own challenges. Interestingly, one can witness a convergence of technologies and investigations across table top and large-scale facilities due to the tremendous prospects for investigation. Suffice it to say that the ability to bring sub-femtosecond, broadband x-ray sources to the table-top has enabled a new era of ultrafast science.

In our article, we will discuss the table top technology and will begin Section I with the state of the art in the generation of attosecond pulses in the soft x-ray (SXR) regime using high-harmonic generation (HHG). We will then discuss the investigative techniques, namely x-ray absorption near-edge spectroscopy (XANES) and extended x-ray absorption fine structure (EXAFS) spectroscopy, enabled by the broadband nature of ultrafast HHG sources. Next follows an overview of recent works that investigate the sub-cycle, nonlinear optical response in solid state materials based on absorption spectroscopy. In Section II, we discuss the state of the art which unites isolated attosecond pulse resolution with core-level XANES. A case study is presented in which XANES is applied to the quantum material titanium disulfide ($TiS_2$). This is the first investigation of a semi-metallic[22–24] transition metal dichalcogen (TMDC) with an attosecond SXR probe at the Ti $L_{2,3}$ edge. The intent of this section is to showcase what is currently possible with attosecond XANES in the water-window SXR regime to stimulate interdisciplinary collaborations between the solid-state, materials, ultrafast x-ray and Attoscience communities. The realization of such collaborations will serve to guide the development of attosecond table-top x-ray sources, to further x-ray spectroscopy theoretical methods to better model complicated material systems, and to advance our understanding of complex, multi-component quantum materials of relevance to address today's problems.

### I.A. Attosecond High Harmonic Generation Sources in the Soft X-Ray Regime

Key to the development of table-top x-ray sources has been the advancement of HHG[25], a process by which visible and infrared (IR) laser photons are energy up-converted to wavelengths spanning the extreme ultraviolet (XUV) and soft x-ray (SXR) regimes. Briefly, optical light is focused into a gas medium where valence electrons are ionized as the electric field of the driving pulse reaches its maximum. Each newly freed electron is accelerated away from its parent ion, picking up kinetic energy as it travels along a negative gradient potential energy surface. When the oscillating electric field eventually reverses sign, the electron reverses its direction and can recombine with its parent ion, resulting in a release of the excess kinetic energy it picked up from the laser field as a burst of x-rays. Because the HHG process is fully coherent and preserves the properties of the driving laser, the resulting x-rays exhibit extraordinary spatio-temporal coherence and cover broadband spectral ranges with down to sub-100-attosecond temporal durations[12]. While HHG performed with widely used Ti:Sapphire 800 nm driving pulses has been a highly successful method for producing femtosecond and sub-femtosecond XUV pulses up to 120 eV, the last decade has seen a tremendous advancement in source development, enabling the generation of attosecond x-ray pulses extending well into the SXR regime[9,14]. Here, we discuss these advancements by highlighting the tunable parameters of the HHG process.

For one, the maximum possible energy of up-converted photons is given by

$$E_{cutoff} = I_p + 3.17U_p \tag{1}$$

where $I_p$ is the ionization potential of the gas target, almost always chosen to be a noble gas, and $U_p$ is the quiver energy that scales as $\propto I_L\lambda_L^2$, with $I_L$ and $\lambda_L$, the intensity and wavelength of the driving laser. In initial works, a conventional approach to increasing $E_{cutoff}$ was simply to increase the intensity of the optical driver. This method, however, proves to be limited, as it is desirable to stay within the tunnel ionization regime and avoid plasma effects. Given equation (1), an alternative approach is to increase the optical pulse's wavelength since a slower oscillating electric field gives the emitted electron more time gaining kinetic energy before the field reverses sign. It is, therefore, now common for HHG setups to utilize optical parametric amplifiers (OPA) or OPAs combined with a parametric chirped amplification (OPCPA)[26–28] scheme to down-convert 800-nm photons to IR wavelengths of 1-2-μm. Of course, OPA is a nonlinear process

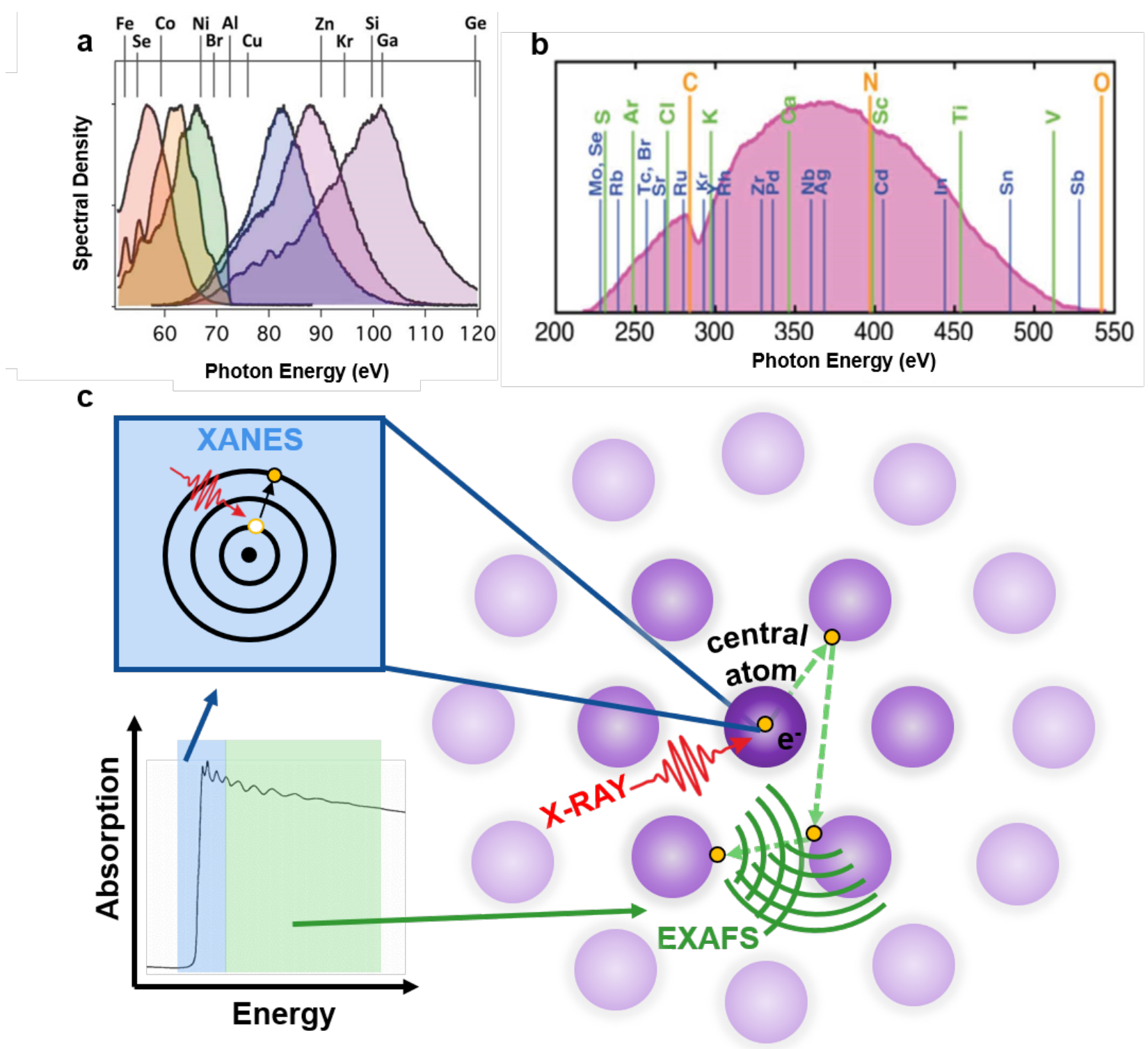

**Fig. 1:** Output energies of HHG-based attosecond sources and schematic of XAFS measurements. **a** Representative isolated attosecond pulse spectra generated with an 800 nm single-cycle driver pulse covering relevant absorption edges the XUV region. **b** A representative spectrum of an isolated attosecond pulse covering over 300 eV in the SXR regime generated with a single-cycle 1800 nm driver pulse. Multiple element edges are covered with K-edges, L-edges, and M-edges in orange, green, and blue, respectively. Panels b with permission from ref. 13 and c with permission from ref. 25. **c** XANES measures unoccupied valence states (VS) using a core-electron excited by a high energy SXR photon and is, therefore, energetically relevant around absorption edges. In time-resolved XANES typically an optical pulse first triggers an out-of-equilibrium dynamic in the valence states. EXAFS spectroscopy instead probes a larger energy window above the edge where modulations in the absorption spectrum stem from photoelectrons with enough excess energy to scatter multiple times and auto-interfere with itself.

characterized by conversion efficiencies of around 30%, so a trade-off is made in terms of intensity. Additionally, the wavelength cannot be increased indefinitely, as the probability of recombination decreases dramatically the further away the electron travels from its parent ion. With carefully balancing the choice of driving wavelength, intensity, and gas, HHG sources have become readily tunable to reach wavelengths spanning the XUV to the SXR regimes (Fig. 1a,b)

It is important to realize that with the realization of sub-femtosecond pulses come extremely broad bandwidths. While such broad bandwidths are unwanted for photoemission spectroscopies due to emission from multiple states, they are a decisive advantage for absorption spectroscopy of multi-component material systems. Simultaneously access to the multiple component absorption edges allows to disentangle lattice, charge, and spin dynamics[30] of all the elemental components of the material at once. Especially the SXR regime is important for such investigations, as compared to the XUV, since elemental edges overlap less and the correspondingly high photon energies allow access to K and L-edges of many first-row transition metals and common material constituents, like S,C,N and O. Attosecond SXR radiation are thus ideal tools for organic chemists, biochemists, materials scientists, and solid-state physicists alike[31] since investigations of composite materials, quantum materials, adsorbates, organic electronics, twisted bilayer graphene[32–34], high-temperature superconductors[35], metal-organic frameworks[36,37], carbon nanomaterials[38,39], NV-diamond centers[40,41], and transition metal dichalcogenides (TMDCs)[22,42] will be possible.

Of the experimental probing schemes one can implement with a table-top x-ray source, x-ray absorption spectroscopy (XAS) is most appealing since it directly leverages the extreme bandwidth of the attosecond pulse. The absorption of light by a sample is directly related to the imaginary part of its dielectric function and XAS is a sensitive probe of element-specific electronic structure when the incident light is resonant with an atomic edge, and if that transition is unique. These conditions are fulfilled with at high photon energies which level core transitions, thus avoiding multiplet effects and excessive photoionization. XAS methods can be further divided into x-ray absorption near-edge spectroscopy (XANES) and extended x-ray absorption fine-structure spectroscopy (EXAFS), where the difference lies in the spectral region around the absorption edge probed[18,43] (Fig. 1c). Without an exact division of energy ranges, XANES roughly covers the region from the edge onset to about 50 eV above, such that absorption is dominated by resonant transitions from occupied core states to unoccupied valence states. In this regime, information about an atom's density of states, site symmetry, oxidation state, and coordination number can be obtained since the photoelectron's mean free path is on the order of the first-coordination shell. On the other hand, EXAFS is characterized by high-energy photoelectrons that travel far enough to encounter neighboring atoms. The so existing auto-interference between the different scattering pathways of a photoelectron will modulate its scattering cross section, therefore result in undulations of the absorption in the EXAFS regime, which can be processed by a Fourier analysis to provide information about lattice spacings and symmetry.

Key to both of these x-ray fine structure spectroscopies (XAFS) is the ability to resolve subtle features in the absorption spectrum, which necessitates high-brightness, stable x-ray sources. For this reason, the development of synchrotron sources was instrumental to the viability of XAFS techniques. With the shortening of pulse durations and the advancement of timing schemes, time-resolved XAFS has now become a popular method for investigating nonequilibrium electronic structure dynamics. Here, the >30 picosecond x-ray pulses from electron storage rings can be shortened to hundreds-of-femtoseconds with "femtoslicing" techniques, but even these pulse durations are too long to resolve electron wavepackets that dephase with a characteristic timescale of femtoseconds[44]. On the other hand, XFELs can offer tens-of-femtosecond pulse durations, but tend to be quite narrowband and spectrally unstable. For XFELs stability, longitudinal coherence, and synchronization are significant challenges that render time-resolved fine-structure studies challenging at the present time.

Attosecond transient absorption spectroscopy (ATAS) is therefore ideally suited to exploit the prospects of time-resolved XAFS techniques to the sub-femtosecond timescales on which electron motion typically occurs. Furthermore, the broadband nature of HHG-based x-ray sources capable of covering hundreds-of-eV in the SXR regime allows for the simultaneous acquisition of XANES and EXAFS in a single shot with $\Delta E/E \approx 1/1000$ spectral resolution. Our group recently demonstrated combined static XAFS spectra covering over 300 eV around the Carbon K-edge in graphite, thus paving the way for attosecond, time-resolved XAFS studies[18]. Indeed, the ultrashort pulse durations offered by HHG sources and their broadband spectral characteristics will be indispensable in extending the toolbox for next-generation ultrafast science.

### I.B. Attosecond Spectroscopy: The Sub-cycle Optoelectronic Response in Solids

We now turn our discussion to the science that has been enabled by table-top attosecond x-ray spectroscopy, with a specific focus on attosecond transient absorption spectroscopy (ATAS) applied to solid-state systems. For a more specified discussion on ultrafast gas phase dynamics, we refer the reader to recent excellent reviews[45,46]. A series of landmark works have focused on questions related to the speed and mechanisms with which optical pulses affect material properties, with motivations driven in large-part by the prospects of light-wave electronics controlled at petahertz (PHz) frequencies. With single-cycle visible light pulses having few-femtosecond pulse durations, attosecond probing pulses are necessary. Further, the elemental-specificity and separation in energy-scale between the optical and XUV pulses have been invaluable to the interpretation of results.

Initial optical pump – XUV probe experiments were performed on $SiO_2$ and pure Si in the strong-field excitation regime ($10^{12}$ W $cm^{-2}$) with near critical electric fields of 2.5 V/Å and sub-band gap pulse energies, conditions under which nonlinear effects would be expected to play a leading role. Indeed, reversible oscillations occurring only during the time-overlap of the pump and probe pulses at twice the pump frequency (2ω) and step-like excitation behavior following the time-trace of the optical pulse were observed, respectively[16,47]. Another experiment on GaN, a 3.35 eV bandgap semiconductor, in the weak-field excitation regime demonstrated photo-excited 3ω dipole oscillations, thus showing PHz field control of carrier dynamics via a three-photon absorption process[48]. The wide bandgap insulator $Al_2O_3$ doped with Cr ions was subsequently studied in a similar experiment, for which it was shown that multi-PHz frequency manipulation was possible via interference of multi-photon processes with frequency tunability offered by choosing the identity of the chemical dopant[49].

While the aforementioned experiments highlight optical control of semiconductors and dielectrics in the strong field regimes, Lucchini *et al.* sought to understand the competition between intra- and inter-band effects at intermediate field strengths[50]. In this regime, the optoelectronic response should be described by a hybrid of the strong- and weak-field extrema: in the weak- (strong-) field limit, inter- (intra) band transitions should dominate with the pump light behaving like a photon (field). Here a 5-fs CEP stabilized below-gap IR pulse was incident on diamond, a dielectric with a 7.3 eV indirect bandgap. Whereas the XUV pulse is most commonly used as a probe via core-to-valence excitations, the 250 as, 42 eV probe pulse excited valence electrons to the CB continuum, the intent here being to more cleanly distinguish between inter- and intra-band effects. The resulting transient absorption spectra revealed 2ω oscillations, which were attributed to the dynamical Franz Keldysh effect (DKFE). The DKFE, in which the electric field of the driving pulse bends the crystal potential enough to distort electronic wavefunctions to leak into the band gap and make possible photon-assisted tunneling, was shown to nearly fully account for the measured signal. This experiment was recently re-performed and found to be almost instantaneous, occurring with a 49 as time constant [51].

Building on these findings, Schlaepfer *et al.* and Volkov *et al.* used XUV-ATAS to investigate inter- and intra-band effects on the sub-cycle optoelectronic response under two different circumstances, namely, when the pump pulse is resonant with the bandgap of a semiconductor and when the material system is metallic with no bandgap at all[52,53]. In the former, it was expected that inter-band carrier dynamics would come to the forefront. Surprisingly, the observed 2ω oscillations were, again, found to be dominated by intra-band currents, though the coupling between inter- and intra-band effects was necessary to explain the enhanced excitation of carriers across the bandgap. In the latter, ATAS was performed at the Ti $M_{2,3}$ edge in pure Ti in response to a $1 \times 10^{12}$ W $cm^{-2}$ NIR pump pulse (10 fs, 1.55 eV). It was anticipated that many-body electron correlation effects (i.e. screening, collective electron motion) from the Ti 3d valence electrons would alter the relative weighting of intra- and inter-band effects in the sub-cycle optoelectronic response. Unlike in the nonzero bandgap materials where oscillatory effects were observed, the time-resolved spectrograms instead depicted a quasi-instantaneous linear response with an optical density transient following the pump laser fluence. This finding was interpreted as a breakdown of the independent-particle approximation, indeed, a consequence of correlation-driven effects. The pump pulse was found to strongly interact with conduction band electrons, causing increased electron localization and therefore strong screening, which was confirmed by theoretical calculations to manifest itself as ultrafast and linearly-dependent on pump fluence.

From the aforementioned results, we highlight several important observations. For one, in less than 10 years, the realization state-of-the-art ATAS beamlines have greatly enhanced our fundamental understanding of the light-matter interaction and have brought us one-step closer to realizing PHz-controlled light-wave electronics. It is clear that the sub-cycle optoelectronic response is highly dependent on the strength of the field, the size of the material bandgap compared to the photon energy of the pump pulse, and the inherent material properties, with intra- and inter-band effects playing nonintuitive relative roles. The multi-variate nature of this problem though leaves many unanswered questions and necessitates a more precise probe of the electronic structure and its dynamics. Secondly, we note that real advance for investigations necessitates probe pulses above about 150 eV: Shallow bound states, or semi-core states in the XUV, are usually dipole-allowed but their small binding energies makes the interpretation of measured dynamics only meaningful if multiplet effects can be factored out. Additionally, near-valence transitions suffer from parasitic photoionization ($\sigma \propto Z^5 \omega_{Xray}{}^{-9/2}$), thus altering the final states to be probed significantly[54]. With higher-energy SXR pulses extending up to hundreds of eV, unambiguous state-selectivity can be achieved which permits a clean interpretation of spectra[55,56]. Consequently, attosecond sources at significantly higher photon energies[10,11,14,57], which are much harder to implement, are highly sought-after for state-selective ATAS spectroscopy, i.e. XANES and EXAFS spectroscopy. Lastly, we notice that the field of attoscience has barely scratched the surface in terms of the complexity of materials capable of being studied. The prospects to bring attosecond temporal resolution together with state and element selective core-level XANES is expected to allow a precise investigation of multi-body and correlation effects which preclude our current understanding of many interesting phenomena in material science, ranging from Mott transitions to exotic new phases in composite quantum materials. From an application point of view, simple thin-films, wafers, and foils (i.e. $SiO_2$, Si, Ti, etc.) have been the topics of most ATAS works for several reasons: 1) a transmission geometry necessitates smooth surface, thin-film samples, for which only a handful of commercially available samples meeting these requirements exist, with few-to-none having greatly novel emergent properties 2) except for the lightest elements, most elements do not have core-states in the XUV regime that has previously been the limit to ATAS, and 3) solid-state attoscience is an entirely new field requiring expertise in ultrafast photonics and custom instrumentation that there may exist a disconnect between the those that develop the photonic means and the materials science or synthetic chemistry communities that make interesting materials. Moreover, most theoretical tools in solid state physics have been developed with the aim to predict static electronic structure. Here, entirely new developments are needed to describe the nonlinear interactions arising from the pump and probe photons with the additional challenge to describe the multi-body state of correlated quantum materials. Nevertheless, with the elemental-specificity x-rays can offer, the study of dynamics in more technologically-relevant, multi-component materials of interest to the broader scientific community is only limited by the energies accessible to HHG x-ray sources and the ability to interpret the information contained in the absorption spectrum to extract electronic, spin and lattice information. Considering these prospects, the recent development of attosecond SXR-HHG and spectroscopies such as XANES is truly exciting.

In what follows, we will present a case study which unites all the discussed requirements for the first time. This discussion on attosecond XANES shows how state and element selectivity is achieved and how it is used to extract the ultrafast, sub-cycle electronic occupation of specific electronic states inside the semi-metallic TMDC $TiS_2$. This represents the first attosecond state-resolved spectroscopy measurement of carrier-band dynamics in a TMDC at the Ti $L_{2,3}$ edge and a novel, previously unexplored, perspective on the ultrafast, sub-cycle optoelectronic dynamics that ensue in a semi-metal system in the weak field regime.

## II. A Case Study: Soft X-Ray attosecond XANES to Investigate the Weak-Field Optoelectronic response in the TMDC Semi-metal $TiS_2$

The exquisite electronic and optical properties of TMDCs stem from their partially-filled d-orbitals resulting in complex phase diagrams and giving rise to correlated carrier dynamics that could be exploited for revolutionary new devices[58,59] in information processing, energy harvesting, or high energy density storage[22,23,60,61]. TMDCs have been unexplored in ATAS in part because of the difficulties to disentangle the involvement of electronic states dependent on the material's elemental constituents, and the out-of-equilibrium response of carriers that arises from the complex energetic landscape of valence and conduction band states. To further the understanding of the behavior of these materials in optical control fields and leverage their properties for future light-field-driven devices, it is required to resolve the microscopic response of particular orbitals on the sub-optical-cycle timescale, which requires attosecond-duration and state-selective probes[62–64] for XANES.

The chosen material, $TiS_2$, exhibits attractive structural and electronic properties because of its semi-metal character; e.g. it has extremely high electron and ion mobility, and its electronic and optical properties are reminiscent of its 2D isomorph[65,66]. The properties of $TiS_2$ arise from a tri-layered structure, shown in Fig. 2a, which consists of hexagonal sheets of cationic $Ti^{4+}(3d^0)$ atoms sandwiched between sheets of S atoms. The atoms within the tri-layers (S-Ti-S) are covalently bonded while the tri-layers are coupled by Van der Waals forces. This weak interlayer bonding results in physical properties that are anisotropic to an extent that the material can be regarded as quasi-two-dimensional solid and thus highly amenable to ion-intercalation in Li-ion batteries[66]. Moreover, the material's band structure changes only marginally when reducing dimensionality from the bulk to the monolayer[67]. These features render $TiS_2$ of interest as a platform for ultrafast optoelectronic devices and field-effect transistors[67], for solid-state batteries[68] and high-density energy storage[69]. Most notably, we point out that there is a great desire to apply ATAS to the study state-specific carrier dynamics in equally anisotropic systems, e.g. 2D TMDC's, which have garnered significant recent excitement due to the discovery of their unique properties (i.e. Moire excitons[70–74], charge density waves[75], superconductivity[76], valley-selective optical excitation[77], magnetism[78], etc.) that are highly tunable with external stimuli and not well-understood. Understanding the carrier dynamics in a 3D TMDC like $TiS_2$ that exhibits a minimal change in band structure with reduced dimensionality also will also serve as a control case to provide insight into edges to probe and properties to look for in future studies performed in the 2D limit.

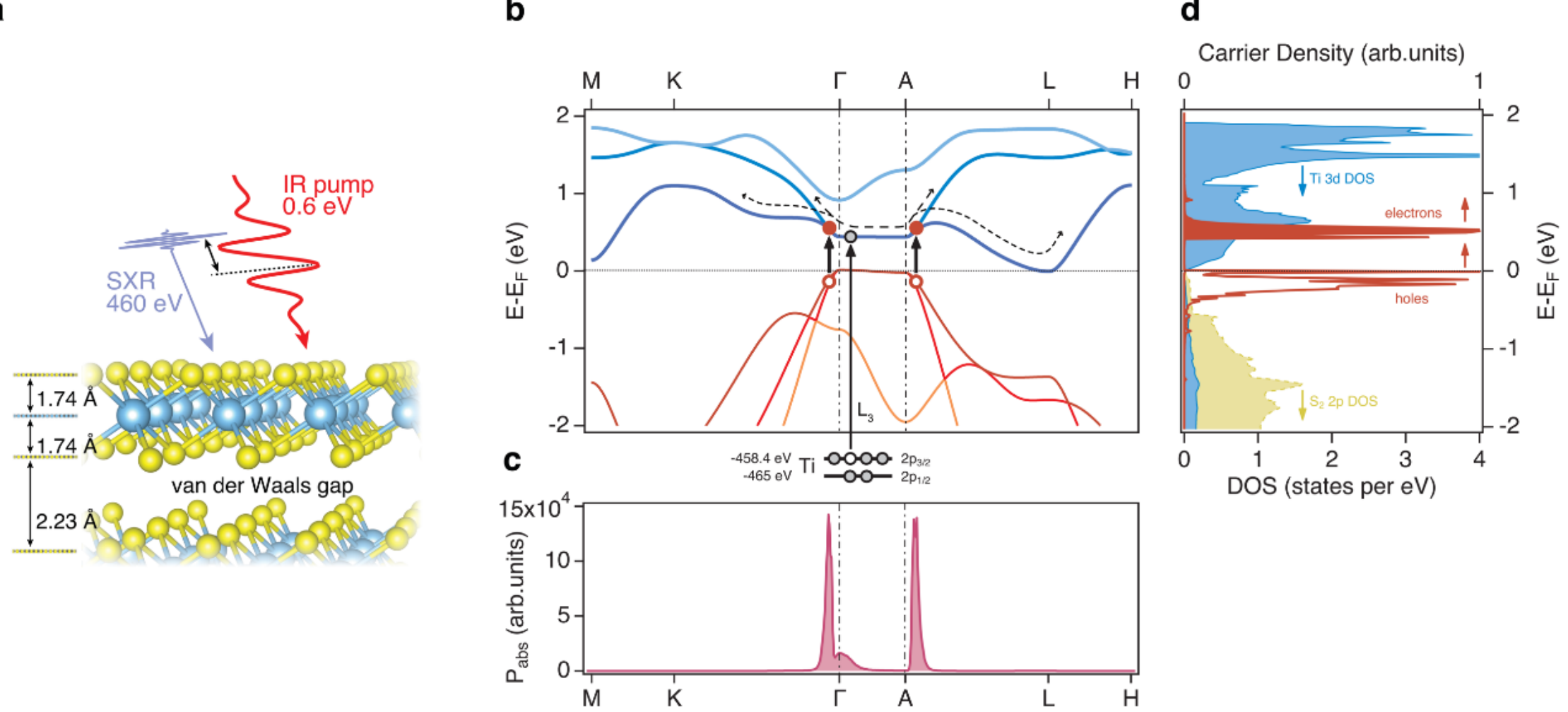

**Fig. 2.** Soft X-ray absorption spectroscopy in $TiS_2$. **a** A sub-2-cycle 1.85-μm (0.6 eV) pump pulse stimulates carrier dynamics inside a 150-nm-thin $TiS_2$ crystal (sketched are two layers), which is probed by a 165-as-duration SXR pulse. **b** DFT calculated band structure showing three valence and conduction bands around the Fermi energy in the first Brillouin zone. Superimposed are single-photon transition pathways between valence and conduction bands from the expected carrier population, shown in **c**, for the experimental excitation spectrum. The excitation is localized around the Γ and A valley points. **d** shows excited carriers depending on energy integrated over momentum space. This corresponds to the electron (red, filled) and hole (red, empty) population, as it would be observed by Pauli blocking in non-momentum resolved XANES, and by assuming the experimental pump spectrum and one-photon transitions. Also shown is the density of states for Ti-3d (blue) and $S_2$-2p (yellow) orbitals. It can be seen that the majority of carrier density is expected in a narrow energy range centered around 0.55 eV above the Fermi energy.

We will show that attosecond-resolution X-ray absorption near edge spectroscopy (XANES[43,62,79,80]), a variation of ATAS that focuses on the near-edge structure, achieves state-selective interrogation of the 3d orbitals of the transition metal atom Ti. Since XANES relies on dipole transitions from core-states, this translates into accessing the 2p-core to 3d-valence transition in Ti and requiring attosecond pulses at a photon energy of 460 eV. By examining the optoelectronic response under previously unexplored conditions, we show that semi-metals can be a viable material for light-wave electronics in the more technologically-relevant case of a weak optical field.

## II.A. Results: Attosecond XANES on $TiS_2$

Here, we demonstrate a state- and attosecond-resolved investigation of the non-equilibrium dynamics of carriers in the 3d-character conduction band in the presence of an electric control field that is weak enough to be realistically achievable in devices, e.g. with plasmonic nano-focusing[81,82], in contrast to the strong-field measurements mentioned earlier. Figure 2b shows the semi-metallic band structure of bulk $TiS_2$ calculated from first principles in the framework of density-functional theory (DFT); see SI. The angular momentum character of the valence bands originates from mixed sulfur 3p and 3s states, while the conduction bands are nearly-exclusively of Ti 3d character.

The opto-electronic response of the material was probed by applying a near-infrared (NIR) optical control field at a photon energy centered at 0.6 eV. This optical field could, thus, directly bridge the 0.23 eV gap of the valence and conduction bands between the Γ and A points by single-photon excitation whilst avoiding excitation of high-lying conduction bands or excursions beyond the first Brillouin zone. Calculations using the spectrum of the optical control field (see SI) show that the excitation is localized around the Γ and A critical points (Fig. 2c). Projecting the momentum-dependent valence excitation onto the transition energy (Fig. 2d) indicates energetically narrow contributions by Pauli blocking to the XANES.

The control field consists of a carrier-envelope-phase (CEP) stable, 1.8-cycle-duration (12 fs FWHM) laser pulse at a center wavelength of 1850 nm, a low-energy replica of the pulse that produced the isolated attosecond SXR pulse through high harmonic generation. The peak intensity of the control field was $4.1 \pm 0.8\ 10^{11}$ W $cm^{-2}$, corresponding to an electric field amplitude of 0.08 V/Å (compare to the 2-2.5 V/Å fields implemented in experiments mentioned in section I.B.) inside the material and an excitation of 0.03 electrons per unit cell, or a carrier density of $1.94 \times 10^{21}$ $cm^{-3}$. An important aspect to resolve in the material response on the scale of the electric field waveform of the control field is, in addition to the SXR pulse attosecond duration, a fast core-hole decay of the material[83]. We estimate a core-hole decay of 3 fs for the Ti 2p states from Ref. [84]. The core-hole decay is faster than the cycle period of the control field (6.1 fs at 0.6 eV), which is consistent with our measurement that shows that the dynamics are indeed resolved on the sub-cycle scale of the optical field.

In the experiment, the SXR attosecond probe interrogated a 150-nm-thick, free-standing, mono-crystalline 1T-$TiS_2$ sample at 26° incidence, with respect to the basal plane normal to sample. A 20-µm-thick Ni pinhole of 100-µm diameter was positioned on top of the $TiS_2$ sample to define the area of spatial overlap between the attosecond probe and the IR control field. The measurement was taken with a home-built spectrometer consisting of a reflective 2400 lines/mm grating (Hitachi) and a cooled CCD-camera (PIXIS, Princeton Instruments) for readout. To state-resolve the dynamics of carriers in the 3d-character conduction band with XANES, we use isolated 165-as-duration SXR[14,18,85] pulses with a bandwidth covering 200 to 550 eV, thus accessing the Ti 2p core states at -458.4 eV ($2p_{3/2}$) and -465.5 eV ($2_{p1/2}$); see Fig 3 . Shown in Fig. 3b is the XANES to identify the positions of the Ti $L_2$ and $L_3$ absorption edges that arise due to transitions from the $2p_{1/2}$ and $2p_{3/2}$ core states. The static XANES measurement, i.e. un-pumped and non-time-resolved, serves as reference and it is in excellent agreement with measurements taken at the ALBA synchrotron light source ALBA (Barcelona) and well-reproduced by theory (see SI).

Figure 3a shows the differential absorption spectrum ($\Delta T = T_{pumped} - T_0$) normalized to the unpumped case ($T_0$), as a function of IR-pump/attosecond-SXR-probe delay. White bars represent data points that were sorted out when post-processing data based on signal-to-noise ratio. Negative time values correspond to the SXR probe arriving before the IR control field. A positive (red) value reports an increasing SXR transmission due to the field-induced excitation of the material, thus an increased carrier population in conduction band states. Immediately apparent in the measurement is a transient signal with a maximum amplitude of 10%, at twice the oscillation frequency of the IR optical field ($2\omega_{IR}$)., which exhibits oscillations with excursions to both, positive (red) and negative (blue), values; see the SI for a Fourier analysis. Fig. 3b depicts lineouts of the differential absorption (Fig. 3a) at different time delays and establishes that the control-field-induced carrier dynamics indeed occurs at the bottom of the Ti 3d-character conduction band, closest to the Fermi level. Clearly resolved are the transitions to Ti 3d-character states originating from the two spin-orbit split Ti core states $2p_{1/2}$ and $2p_{3/}$, corresponding to the absorption edges $L_2$ and $L_3$. In the following, we restrict our analysis to the $L_3$ edge to investigate the dynamics of the carriers in Ti 3d orbitals as the signal observed at the $L_2$ edge may be mixed with transitions from the $L_3$ to higher-lying conduction band states.

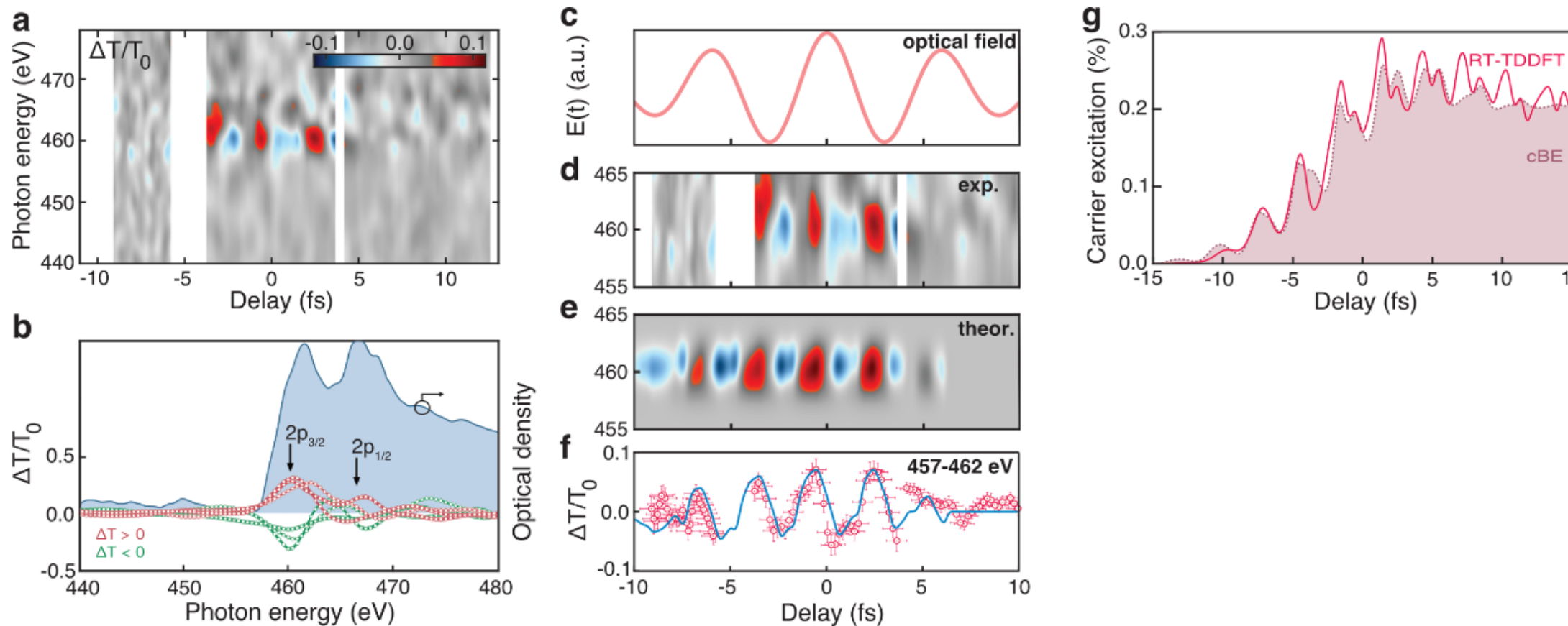

**Fig. 3.** Time-resolved dynamic in $TiS_2$. **a** Differential absorption spectrum as a function of pump-probe delay. ΔT is normalized to the un-pumped case $T_0$ for each time delay. Blanked out areas indicate time ranges without detection. Clearly visible are sub-cycle oscillations of negative and positive differentials with a maximum amplitude of 10%. Time delays were 100 as for negative delays to -5 fs with 6 min integration and 200 as steps with 5 min integration elsewhere. **b** shows spectral lineouts from **a**, taken at the positive and negative extrema (spectral oscillations). The dynamics, red for increased and green for diminished ΔT in relation to the two spin-orbit split states $2p_{3/2}$ and $2p_{1/2}$. **c** indicates the optical control field. **d** displays the relevant subsection of **a** for comparison with the cBE calculated differential absorption in **e**. **f** overlays the measurement with the numerical result over the energy region 457-462 eV; error bars originate from the regression algorithm and take the temporal duration (300 as upper limit), temporal walk-off in the material (1.3 fs) and CEP jitter (85 as) into account. **g** shows the calculated carrier excitation with both models, cBE and RT-TDDFT, of 0.2% corresponding to a carrier density of 1.94 x $10^{21}$ $cm^{-3}$.

## II.B. State-resolved dynamics

To obtain a detailed physical insight into the underlying carrier dynamics in $TiS_2$, we turned to theory. We performed a first-principles electron dynamics simulation of the full pump-probe experiment based on real-time time-dependent density functional theory (RT-TDDFT)[86]. In addition to the ab-initio model, we developed a core-state-resolved Bloch equation model (cBE)[87] to disentangle the various concurrent inter- and intra-band contributions in response to the control field. The cBE model retains the relevant seven bands in three dimensions; it includes the three highest-occupied valence, two-lowest-unoccupied conduction band states together with the Ti-2p core states, and bandgap renormalization (further details in SI)

The result of the theory, including the full inter- and intra-band dynamics, is shown in Figs. 3e,g and 3. The good qualitative agreement between the measurement (Fig. 3a and d) and calculations (Fig. 3e) is depicted in Fig. 3f, showing the differential absorption integrated over an energy interval of 5 eV where the changes are most prominent (between 457 and 462 eV). Indicated in Fig. 3c is the IR control field periodicity which shows that carriers' response (Figs. 3d-f) occurs on the sub-cycle temporal scale with $2\omega_{IR}$ oscillations. We note that the origin of the $2\omega_{IR}$-oscillation has been previously attributed to the DKFE[50,52] and arises due to the ac-Stark effect which is caused by the external pump field that modulates the electronic wave-function in such a way to permit photon-assisted tunneling to the conduction band. From our model, we find that inter-band transitions predominantly occur in the form of one-photon transitions that directly bridge the gap between valence and conduction bands around the Γ and A critical points, as indicated in Fig. 2, and become observable in the transient XANES experiment.

To predict the magnitude of injection of carriers into the conduction band, we used both theoretical models. Shown in Fig. 3g is the result from a projection of the time-dependent wavefunction of the seven-band cBE and from the RT-TDDFT model onto the conduction band state. We find very good agreement amongst the simulations with an excitation value of 0.2% to the 3d-character conduction band, corresponding to a carrier density of $1.94 \times 10^{21}$ $cm^{-3}$.

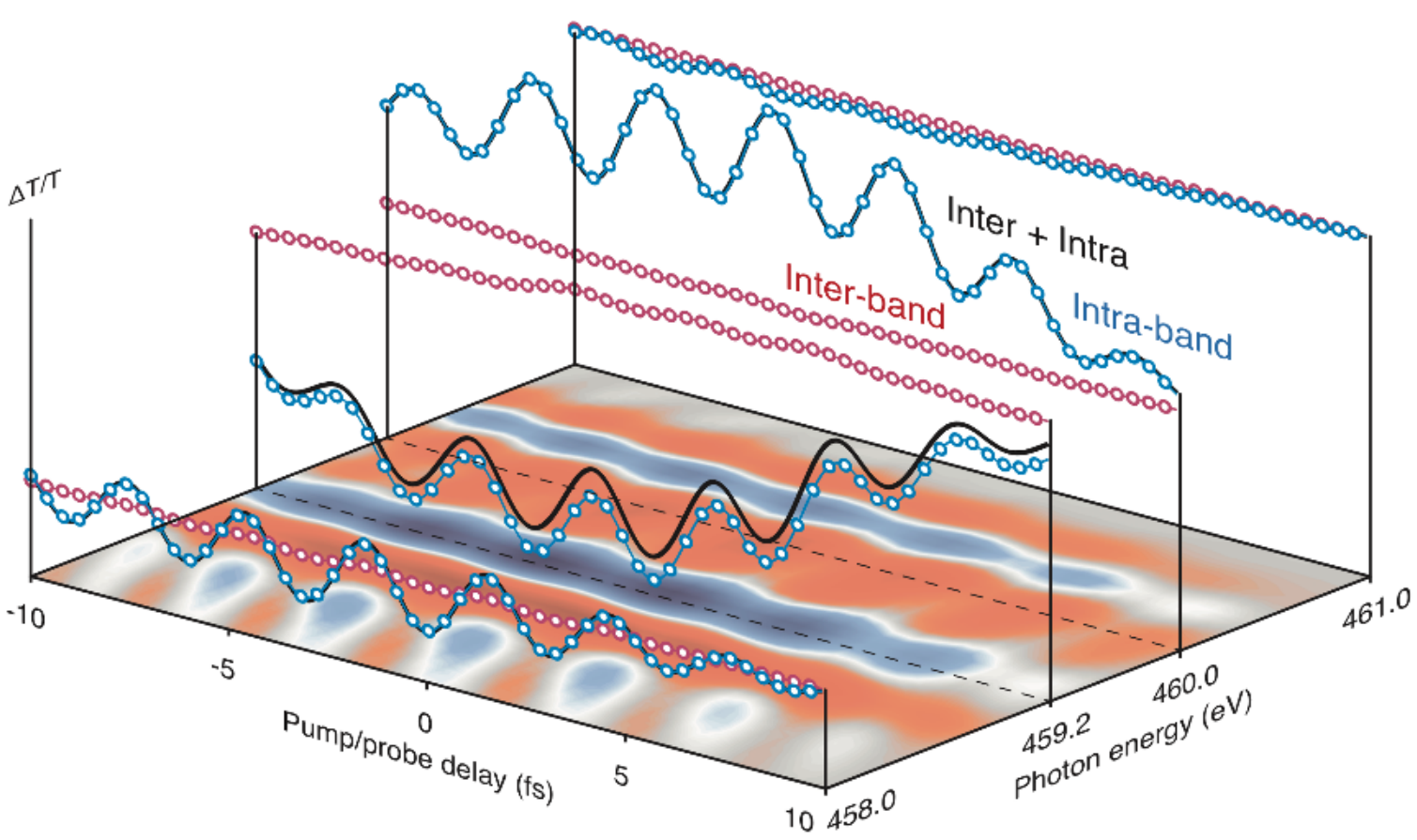

**Fig. 4.** State-resolved carrier dynamics. Shown is the observed material response as function of pump-probe delay, calculated with the cBE model and including 3 VBs and 4 CBs in full 3D. The surface plot displays the full material response ($\log(\Delta T/T_0)$ where red indicates positive and blue indicates negative values). The line-outs show the energy-resolved material response (black curve) which is composed of inter- (red) and intra-band (blue) contributions. Striking is the dominance of intra-band carriers by one order of magnitude over inter-band carriers, they follow the oscillation of the excitation field and modulate the absorption both positively and negatively.

In general, the transfer of electrons to the conduction band at a specific k-space position results in Pauli blocking and, since the core-level transition (Ti-2p to Ti-3d) accesses conduction band states, this leads to bleaching of the SXR absorption. Thus, vital is the role of intra-band motion of carriers as, depending on the mobility, it may diminish (or increase) Pauli blocking at a particular k-space position and manifest as a change of SXR absorption. Therefore, we further investigated the state-resolved dynamics of electrons with a high-resolution cBE calculation with the result shown in Fig. 4. We find that, despite the low field strength, immediately visible is the ubiquitous 2ω-oscillation to both, positive and negative values, arising from both, inter- and intra-band motion. Interestingly, and previously unobserved, the oscillation to both positive and negative values originate predominantly from intra—band currents, owing to the very high carrier mobility of the semimetal and confirming the absence of Rabi cycling. Overlaid as individual lineouts in Fig. 4 is an energy-resolved decomposition of observed carrier motion into inter- and intra-band contributions. Striking is the dominance of intra-band (blue curves) over inter-band (red curves) carriers, and the energy-dependent modulation of the absorption spectrum. The absence of apparent chirp is due to the strong and near instantaneous response of intra-band carriers to the weak driving field. Immediately apparent is the strong modulation of the absorption signatures as function of time. In the SI, we discuss the origin as Fano-line profile modifications that arise during the optical control field due to the time-modified dipole-phase from excitation into the conduction band.

We further find that the intra-band currents stem from accelerating the carrier population only across 38% of the first Brillouin zone in the presence of the weak electric field. It is worth noting that this relatively small spread driven by intra-band currents confirms a weak-field regime in contrast to experiments reported in the realm of strong-field driven solid-state high harmonic generation that regularly drives intra-band currents multiple times across the first Brillouin zone[88].

The intra-band process immediately spreads the conduction band electrons and, thus, reduces the energetic localization of the resonant inter-band excitation, reducing the contrast in the experiment for observing carriers following the localized excitation. This finding is quite nonintuitive because from the semi-metallic band structure (Fig. 2b) one would expect a tunnel barrier opening around the valence band A point and conduction band L point within the first Brillouin zone with minimal external electric field. This is a striking finding for optoelectronic applications of semi-metals and shows that despite gap energies in $TiS_2$ of around 0.23 - 4 eV and a Keldysh parameter of 2.5, tunneling excitation is neither the only, nor the dominant mechanism at play. The material's response is instead governed by predominantly one photon transitions similarly to a semiconductor or dielectric material at much higher field strengths[52,89]. A possible reason for this unexpected behavior of the material in a NIR control field is the high density of states between the Γ and A critical points in conjunction with the very high carrier mobility.

Having investigated the predominant contributions of carrier dynamics to the absorption spectrum, we can exploit the state-selectivity which we have achieved in our measurement to visualize the spatio-temporal flow of carriers based on the theoretical models. Figure 5 shows the time-dependent charge density oscillations and the DOS for two field strengths. Depicted is the change of the electron density distribution that display only a change with respect to the ground state (un-pumped case), with red (blue) isosurfaces indicating increasing (decreasing) density; the anisotropic distribution of this charge-density difference reflects the hybridization between Ti-d and S-p states. We observed a clear charge depletion along the bond and find an associated increase of electronic density on the Ti state with d-character, consistent with the findings of Volkov et al.[53] that demonstrate optically-induced electron localization around Ti atoms. The charge re-distribution occurs within a femtosecond over a distance nearing half the distance between Ti and S atoms of 2.4 Å. The charge re-distribution increased in distance with increasing field amplitude of the NIR laser pulse, thus visualizing the near-infrared field-driven motion of charged carriers within the unit cell. Interestingly, according to the RT-TDDFT computations at the low field amplitude of 0.08 V/Å, used in this study, the excitation profile of carriers promoted into the conduction band density of states (DOS) mimics the pump pulse photon energy spectrum with transitions originating from valence band states just below the Fermi energy into conduction band states approx. 0.5-1 eV above the Fermi energy; note this is in very good agreement with

Fig. 2d. Clearly discernible is also an energy gap between the depopulated valence and populated conduction band states.

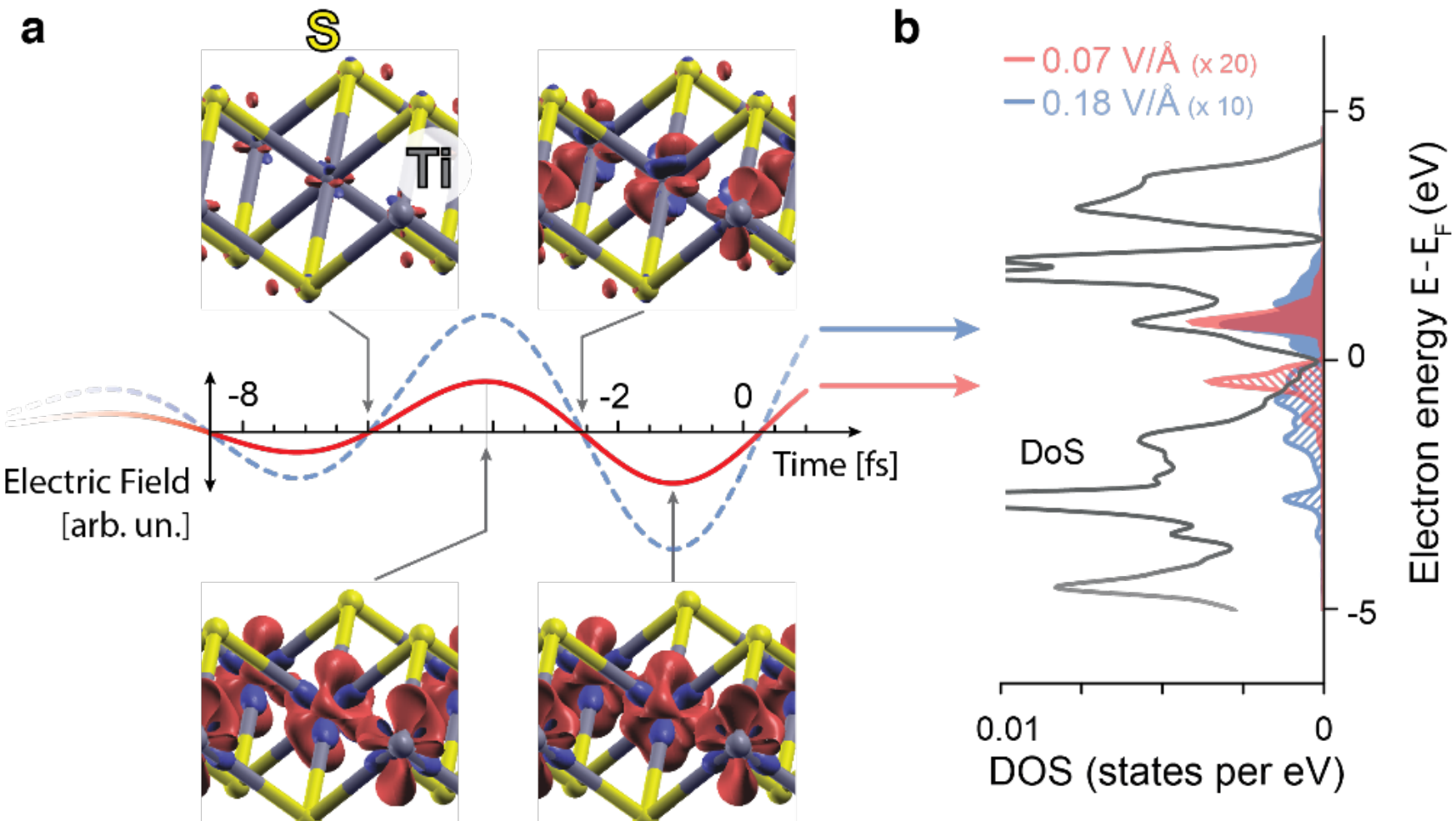

**Fig. 5.** Spatio-temporal carrier dynamics. **a**: At different moments within the electric field oscillation, the difference of electron density is displayed with respect to the ground state (4 panels showing the carrier density displacement within the unit cell). Increased density is shown in red and decreasing density in blue. The time-dependent difference shows enhanced localization of population at Ti-d orbitals and these changes occur with a spatial extent up to half the nearest neighbour distance and oscillate predominately with the half-cycle period. **b**: Density of states (DOS) and light-induced occupation: RT-TDDFT computations observe the transition from a distribution of occupied states within the DOS split in accordance with the pump pulse spectrum for the field amplitude studied in this experiment (occupied DOS in red) towards a more distributed excitation profile extending both into higher and lower energetic intervals of the valence and conduction bands already for a peak field amplitude approximately twice the experimental value (occupied DOS in blue).

The present measurement provides the first unambiguous investigation of the Ti:3d binding orbital which determines the response of the TMDC to an external light field. Our investigation reveals that localization of electron density occurs around the Ti atoms within a fraction of the single-cycle of the optical control field. The change in electronic density is strikingly large for the weak optical excitation field and it modifies the semi-metallic TMDC, leading to strong intra-band currents and a behavior alike the DKFE which was previously only observed in a wide-bandgap system[50]. This raises the interesting question as to whether the DKFE is truly a strong-field effect, and we note that the relevant metric is rather the strength of the control field relative to the size of the bandgap. If the latter is true, semi-metals will be promising candidates for light-wave electronics, requiring much weaker control fields.

## III. Conclusions and Outlook

We have described the recent developments of attosecond technology and x-ray spectroscopy and how the convergence of techniques allows nowadays to probe the fine structure of an absorption edge in a multi-component material system in real time, thus elucidating the state-resolved dynamics of charge carriers in situ. Beyond the background of the field, as an example, we discussed the state of the art with a XANES measurement at the Ti $L_{2,3}$ edge at 460 eV of a TMDC quantum material. We described how the core-state-resolved XANES method allows to isolate the spectral contributions from Ti 3d orbitals, thus accessing the opto-electronic response of the semi-metallic TMDC $TiS_2$ with attosecond temporal resolution and orbital selectivity. The results presented here suggest that $TiS_2$ is a promising candidate for PHz field effect devices since very moderate optical fields achieve efficient photo-doping to strikingly high carrier concentrations and such optical fields are able to modulate the material response quasi-instantaneously, i.e. without measurable delay, on the sub-cycle scale of an optical field. Our investigation of the TMDC $TiS_2$ provides new insight for the usage of semimetal TMDCs as building blocks, such as in optical field effect transistors for the next generation of opto-electronic devices and light-wave electronics. Whilst showcasing what is nowadays possible with attosecond soft x-ray XANES on the particular example of the TMDC, such investigations will allow to study the exact interaction of electrons, holes and lattice modes, thus giving access to the coherent dynamics and its dephasing in a plethora of materials. The prospects are clearly spectacular since the method is applicable to gases, liquids, solids and amorphous systems. Monolayer systems such as graphene, BN or TMDCs can be probed in reflection, while nanometer-size systems such as composite quantum materials like BN-sandwiched graphene and TMDCs, or adsorbates, can be measured in transmission. We note that an insightful connection between the measured absorption spectrum and the embedded electronic and lattice dynamics is only possible in the simplest of systems without theory. New theories are therefore needed, which allow to describe the non-equilibrium dynamics of the multi-body state with the pump and probe photons. Presently, such developments are already under way (largely) based on real-time implementations of density functional theory, either in real or reciprocal space. At the same time, the development of XFEL sources, i.e. the expected improvement on their attosecond temporal resolution and reduced timing jitter, will increase present capabilities to the hard x-ray regime and attract further investigations. Clearly, the possibilities are excellent to gain unprecedented, and comprehensive, insight into the intricate interplay between charge carriers and lattice which determine the response of materials to an external stimulus. We believe that the described tools will become decisive to move towards an engineered approach to design advanced materials and devices based on exact knowledge of their functionalities. While serendipity has served us surprisingly well, e.g. with the recent discovery of twisttronics with magic-angle graphene, today's societal challenges demand drastically new solutions that are tractable in a systematic and predictable way. For instance, the projected increased demand in energy-efficient information processing and storage devices supersedes the growth in energy production over the next 20 years. With the possibility to scrutinize loss channels in electronic transport and material excitation, we would have a tool to identify possible mitigation strategies to approach the long-sought Landauer limit of information processing devices. We envision studies of electron and lattice dynamics on the building blocks of heterostructures based on today's toolbox of conventional and quantum materials such as e.g. Si, Ge, graphene, graphite, BN, TMDCs, and high-Tc superconductors like YBaCuO, TiBaBaCuO or $K_3C_{60}$. The dynamic behavior of these materials may be sufficient for computer models to predict how combinations of such materials can achieve material systems in which e.g. ballistic transport is topologically protected, or in which the valley and spin degrees of freedom are leveraged to store and switch information with THz to PHz speeds. Measurements on the such designed systems will feed back to advance the design. Intimately related with such goal is the investigation of canonical physics questions since they have immediate impact to society. For instance, it is still debated what exactly triggers the superconductive state of a high - Tc superconductor like LaBaCuO, YBaCuO, or e.g. $K_3C_{60}$, and contradicting theories exist to describe the mechanism. New insights into Mott physics would be directly relevant for multi-body quantum dynamics and further our understanding of quantum simulators. As such, the application of ATAS towards the study of materials will likely be of great benefit to re-examine long-standing questions of, for example, phase-transitions[90,91], superconductivity in layered materials[92], and emergent electronic properties in low-dimensional systems.

**Supplementary Contributions**

See the supplementary material for further experimental details such as the static XANES, calibration measurements at the synchrotron light source ALBA, investigation of the pump absorption and dynamics, the data analysis, and information about the sample growth and preparation. We also include information about the various density-functional theory simulations and the core-resolved Bloch Equation Model.

**Author's Contributions**

J.B. conceived and supervised the project. B.B. and I.L. performed the experiments and analyzed the data. N.D.P. and S.C. contributed to the experimental setup. A.P. performed the SBE calculations, C.C., C.D., M.U. and K.Y. conducted the DFT and TDDFT modelling. E.P. and J.H.M. performed calibration measurements at ALBA. S.M.V. and E.C. grew the samples, T.D. cut the samples. IR absorption measurements were conducted with S.W. M.S., E.B. and M.Z. aided J.B. and B.B. in writing the manuscript.

**Acknowledgements**

J.B. and group acknowledge financial support from the European Research Council for ERC Advanced Grant "TRANSFORMER" (788218), ERC Proof of Concept Grant "miniX" (840010), FET-OPEN "PETACom" (829153), FET-OPEN "OPTOlogic" (899794), Laserlab-Europe (EU-H2020 654148), MINECO for Plan Nacional FIS2017-89536-P; AGAUR for 2017 SGR 1639, MINECO for "Severo Ochoa" (SEV- 2015-0522), Fundació Cellex Barcelona, CERCA Programme / Generalitat de Catalunya, the Alexander von Humboldt Foundation for the Friedrich Wilhelm Bessel Prize. A.P. acknowledges Marie Sklodowska-Curie Grant Agreement No. 702565. I.L. acknowledges MINECO for a Juan de la Cierva postdoctoral fellowship and B.B. acknowledges support by the Severo Ochoa Fellowship program for her PhD fellowship. S.M.V. and E.C. acknowledge funding from the EU (Advanced ERC grant SPINMOL), the Spanish MINECO (grant MAT2014-56143-R and Unit of Excellence Maria de Maeztu MDM-2015-0538 and SOMM Alliance); S.M.V thanks the Spanish MINECO for FPU grant (FPU14/04407). M.U. was supported in part by MEXT as a social and scientific priority issue (Creation of new functional devices and high-performance materials to support next-generation industries) to be tackled by using post-K computer. M.Z. acknowledges support by the Max Planck Society (Max Planck Research Group) and the Federal Ministry of Education and Research (BMBF) under "Make our Planet Great Again – German Research Initiative" (Grant No. 57427209 "QUESTforENERGY") implemented by DAAD. C.D. and C.C. acknowledge partial support from the German Research Foundation (DFG) and IRIS Adlershof. We thank Dr. D. Zalvidea, Prof. A. Bachthold, Prof. J. Garcia de Abajo, H.-T. Chang for helpful and inspiring discussions and Prof. T. Ra'anan, Prof. C. Ropers and Dr. P Schmidt for assistance with the sample preparation.